\documentclass[a4, 11pt]{article}
\usepackage{amsmath} %Never write a paper without using amsmath for its many new commands
\usepackage{amssymb} %Some extra symbols
\usepackage{geometry} % to change the page dimensions
\usepackage{graphicx} % support the \includegraphics command and options
\usepackage{epsfig}
\usepackage{booktabs} % for much better looking tables
\usepackage{array} % for better arrays (eg matrices) in maths
\usepackage{paralist} % very flexible & customisable lists (eg. enumerate/itemize, etc.)
\usepackage{verbatim} % adds environment for commenting out blocks of text & for better verbatim
\usepackage{subfig} % make it possible to include more than one captioned figure/table in a single float
\usepackage[dutch]{babel}
\usepackage{fancyhdr} % This should be set AFTER setting up the page geometry
\usepackage{sectsty}
\allsectionsfont{\sffamily\mdseries\upshape} % (See the fntguide.pdf for font help)
\usepackage[titles,subfigure]{tocloft} % Alter the style of the Table of Contents

\title{Effcient Resource Allocation through Integer Linear Programming: a detailed example}
\author{Filip De Turck, Ghent University - imec, Belgium}   
\date{}
\begin{document}
\maketitle
\renewcommand{\abstractname}{Abstract}
\renewcommand{\refname}{References}
\begin{abstract}
In this paper, we show how a resource allocation problem can be solved through Integer Linear Programming (ILP). A detailed illustrative example is presented, together with an exhaustive overview of the mathematical model. The size of the required vectors and matrix are determined as well. The presented example can be used to learn students the fundamental basics of ILP-based resource allocation. Next, the specific benefits of the ILP approach compared to other resource allocation algorithms are outlined in this paper. Finally, a related work section is provided with relevant references for further reading. The provided references contain examples of ILP-based resource allocation in modern networks and computing infrastructures.
\end{abstract}

\section{Introduction and example}
As an illuustrative example of resource allocation, we consider a cluster of 5 computers, where we want to allocate resources for 10 simultaneously running software components. The resource allocation should be done in the most energy efficient way . The software components are characterized by the required CPU cycles/s and memory requirements. In the sections below, the resource allocation problem will be formulated as an Integer Linear Program (ILP) by means of a mathematical model. Next, the size of the vectors and matrix in the example ILP formulation will be determined in section~\ref{section:size_calculation}.

\section{Mathematical model}
In an Integer Linear Programming (ILP) based approach, we need to formulate the resource assignment problem as follows:\\\\
\texttt{maximize \textbf{$c^T x$}\\
subject to \textbf{$A x \leq b$}, \textbf{$x \geq 0$}.\\
where the values of the \textbf{$x$} vector are integer.}\\\\\\
In other words, we need to determine the values of the vector \textbf{$c$} (N elements), the matrix \textbf{$A$} (M rows and N columns), and the vector \textbf{$b$} (M elements).
The vector \textbf{$x$} contains the decision variables (N elements).\\
Note that \textbf{$c^T$} denotes the transposed vector \textbf{$c$} to allow multiplication with the \textbf{$x$} vector.\\
The vectors \textbf{$b$} and \textbf{$c$} and the matrix \textbf{$A$} can be entirely filled out based on the input parameters, i.e. all numerical values for these two vectors and matrix can be determined.
Next, we can input them to an ILP solver (e.g. CPLEX), which will generate the optimal values for the \textbf{$x$} vector.\\
Based on these \textbf{$x$} vector values, the optimal resource allocation can be done.

\section{Determination of the matrix and the vectors}
\subsection{Decision variables}
The following decisions need to be made:
\begin{itemize}
	\item for each software component, on which computer to start the component,
	\item for each computer, whether or not to switch this computer on.
\end{itemize}
We introduce the following binary variables (which can take the values 0 or 1):
\begin{itemize}
	\item $d_{ij}$ for all i (0 ... 9) and j (0 ... 4),
	\item $o_j$ for all j (0 ... 4).
\end{itemize}
When software component $i$ is started on computer $j$, $d_{ij}$ equals 1, and 0 otherwise.\\
Similarly, when computer $j$ is switched on, $o_j$ equals 1, and 0 otherwise.\\\\
The \textbf{$x$} vector contains the $d_{ij}$ and $o_j$ variables (10$\times$5+5=55 variables in total for this assignment), put one after the other in the \textbf{$x$} vector.\\

\subsection{Vector \textbf{$c$}}
The objective of the Integer Linear Program is to maximize \textbf{$c^T x$}, i.e. the values of the \textbf{$c$} vector are multiplied one by one with the decision variables in the \textbf{$x$} vector and summed.\\
As stated in the assignment, the objective is a resource allocation in the most energy efficient way, i.e. switch off as many computers as possible.
In other words, we need to minimize:\\
\begin{equation}
\sum_{j=0}^{4}o_j
\end{equation}\\
this comes down to maximize:\\
\begin{equation}
-\sum_{j=0}^{4}o_j
\end{equation}\\

in other words, the values -1 are inserted in the \textbf{$c$} vector at the corresponding places (5 in total) and the other elements get value 0 (50 in total). In case we want to favour some computers to get switched off if possible, we can give them a value lower than -1 in the \textbf{$c$} vector.

\subsection{Matrix \textbf{$A$} and vector \textbf{$b$}}
This matrix and vector are needed to express the constraints. The following constraints can be distinguished:
\paragraph{Capacity constraints}
The required memory for the allocated components on one machine can not exceed the available memory on that machine, i.e. for all machines $j$:\\
\begin{equation}
\sum_{i=0}^{9}d_{ij}\times \text{mem}_i \leq \text{mem}_j \times o_j
\end{equation}\\
$\text{mem}_i$ denotes the required memory for software component $i$ and $\text{mem}_j$ the available memory on computer $j$. When $o_j$ equals zero, no software components can be allocated to computer $j$, and all $d_{ij}$ values for this particular computer $j$ need to be zero. When the sum is zero, indeed all terms of the sum need to be zero.\\
In total, this results in 5 constraints for this assignment. Each constraint results in a row in the matrix \textbf{$A$}, where the row values are determined by the above expression. Similarly, it results in 5 entries in the vector \textbf{$b$}.\\\\
Similarly, the constraints for the CPU capacity can be expressed. The required CPU capacity for the allocated components on one machine can not exceed the available CPU capacity on that machine, i.e. for all machines $j$:\\
\begin{equation}
\sum_{i=0}^{9}d_{ij}\times \text{CPU}_i \leq \text{CPU}_j \times o_j
\end{equation}\\
$\text{CPU}_i$ denotes the required CPU cycles for software component $i$ and $\text{CPU}_j$ the available CPU capacity on computer $j$. Similar to the previous constraints, when $o_j$ equals zero no software components can be allocated to computer $j$.\\
In total, this also results in 5 constraints for this assignment, each corresponding to a row in the matrix \textbf{$A$} and 5 entries in the vector \textbf{$b$}.
 
\paragraph{Binary variables}
The decision variables $d_{ij}$ and $o_j$ are binary, which means: for all software components $i$ and computers $j$:\\
\begin{equation}
d_{ij} \leq 1, 
\end{equation}\\
and for all computers $j$:\\
\begin{equation}
o_{j} \leq 1. 
\end{equation}\\
In total, this results in 55 constraints for this assignment. Each constraint results in a row in the matrix \textbf{$A$}, where for each row all values are 0, except for one row element, where the value is 1. Each constraint also results in an entry with value 1 in the vector \textbf{$b$}.

\paragraph{Exactly one allocation} We also need to specify that each software component should be allocated to exactly one computer, i.e. for all software components $i$:
\begin{equation}
\sum_{j=0}^{4}d_{ij} = 1. 
\end{equation}\\
In total, this results in 10 constraints for this assignment and 10 corresponding rows in the matrix \textbf{$A$}, together with 10 entries with value 1 in the vector \textbf{$b$}. These constraints will force some $d_{ij}$ values to $1$ and equations (3) and (4) will then force some $o_j$ values to $1$.\\\\

In conclusion, the expressions above allow us to fill out the matrix \textbf{$A$} and the vector \textbf{$b$}. As can be understood from the construction of the matrix \textbf{$A$}, it will be a sparse matrix for this assignment, with many zero values and a relatively small number of non-zero values.

\section{Calculating the size of the matrix}
\label{section:size_calculation}
Given the formulation in the previous section, we can calculate M (number of rows in \textbf{$A$} matrix, and size of \textbf{$b$} vector) and N (size of \textbf{$x$} and \textbf{$c$} vectors, and number of columns in \textbf{$A$} matrix) for this assignment as follows:\\\\
\texttt{M = 5 + 5 + 5$\times$10 + 5 + 10 = 75,\\\\
N = 5$\times$10 + 5 = 55.}\\

\section{Specific benefits of ILP compared to other resource allocation algorithms}
When an optimization problem can be formulated as an Integer Linear Program, the optimal solution can be calculated by means of an ILP solver, i.e. no other algorithm can determine an \textbf{$x$} vector with integer values, which will result in a higher \textbf{$c^T x$} value, while respecting the constraints \textbf{$A x \leq b$} and \textbf{$x \geq 0$}.\\
The computational time for the ILP solver to determine the optimal \textbf{$x$} vector values can be high, especially when there are a large number of variables and constraints (i.e. large values of N and M). In this case, heuristical algorithms will in general result in faster computation times, but the obtained \textbf{$c^T x$} value will be lower or equal to the \textbf{$c^T x$} value obtained by an ILP solver. There is clearly a trade-off between optimality and calculation speed.\\Typically, when the values of M or N exceed 20000, an ILP solver might need several hours or even days to determine the optimal \textbf{$x$} vector values, depending on the specific problem instances.

\section{Related work}  
ILP-based based resource allocation algorithms and their evaluation in the context of softwarized network management have been published in \cite{vnfp} and \cite{tnsmmoens}. We also refer to \cite{noms2014moens} for ILP-based resource management in hierarchical clouds and to \cite{JNSMmoens} for multi-tenant cloud management. Efficient resource management for virtual desktop cloud computing has been succesfully addressed in \cite{SupercomputingDeboosere} by means of Integer Linear Programming. Furthermore, the technique has been used to optimize the delivery of adaptive video streaming services, as reported upon in \cite{ToMBouten} and \cite{ComNetBouten}, and replica placement in ring based content delivery networks~\cite{ComComWauters}. In addition, ILP-based resource allocation algorithms for Smart Cities \cite{EntropySantos} have been published in~\cite{CNSM2017Santos}.

\end{document}